\documentclass[reprint,amsmath,amssymb,pra,aps,nofootinbib,superscriptaddress,floatfix]{revtex4-1}

\usepackage[T1]{fontenc}
\usepackage[utf8]{inputenc}
\usepackage{amsmath, amssymb}
\usepackage{bbold}
\usepackage{bm}
\usepackage{subfigure}
\usepackage[dvipsnames]{xcolor}
\definecolor{dcyan}{RGB}{0,100,100}
\usepackage{graphicx}
\usepackage{float}
\usepackage{physics}
\usepackage[sectionbib]{bibunits}
\defaultbibliographystyle{naturemag}
\defaultbibliography{references}
\usepackage{hyperref}
\usepackage[capitalise]{cleveref}
\definecolor{green_cust}{RGB}{0,154,85}
\definecolor{red_cust}{RGB}{173,49,54}
\definecolor{blue_cust}{RGB}{0,103,148}
\hypersetup{colorlinks=true, linkcolor=red_cust, citecolor=green_cust, filecolor=blue_cust, urlcolor=blue_cust}

\makeatletter 
    
\renewcommand\onecolumngrid{% <<<<<<
\do@columngrid{one}{\@ne}%
\def\set@footnotewidth{\onecolumngrid}% <<<<<<<<<<<<<<<<
\def\footnoterule{\kern-6pt\hrule width 1.5in\kern6pt}%
}

\renewcommand\twocolumngrid{% <<<<<<
        \def\footnoterule{% restore rule
        \dimen@\skip\footins\divide\dimen@\thr@@
        \kern-\dimen@\hrule width.5in\kern\dimen@}
        \do@columngrid{mlt}{\tw@}
}%

\makeatother    
\usepackage{soul}

\newcommand{\Figref}[1]{Fig.~\hyperref[#1]{\ref{#1}}}

\newcommand{\paird}{44 D_{\frac{5}{2},\frac{5}{2}}}
\usepackage{xspace}
\newcommand{\gtwo}{g^{(2)}(\tau)\xspace}
\newcommand{\cthree}{$C_3$\xspace}
\newcommand{\csix}{$C_6$\xspace}
\newcommand{\foerst}{F{\"o}rster\xspace}
\newcommand{\gtwof}{\ensuremath{g^{(2)}(0) = 0.38 (1)}\xspace}
\newcommand{\gtwon}{\ensuremath{g^{(2)}(0) = 1.0 (1)}\xspace}

\begin{document}

\title{Enhanced Rydberg Blockade through RF-tuned \foerst Resonance}
\author{Lukas Palm}
\affiliation{James Franck Institute and Department of Physics, The University of Chicago, Chicago, IL}
\affiliation{Department of Applied Physics, Stanford University, Stanford, CA}
\author{Bowen Li} 
\affiliation{Department of Physics, Stanford University, Stanford, CA}
\author{Yiming Cady Feng}
\affiliation{Department of Applied Physics, Stanford University, Stanford, CA}
\author{Marius J\"urgensen}
\affiliation{Department of Physics, Stanford University, Stanford, CA}
\author{Jon Simon}
\affiliation{Department of Applied Physics, Stanford University, Stanford, CA}
\affiliation{Department of Physics, Stanford University, Stanford, CA}

\date{\today}

\begin{bibunit}

\begin{abstract}
Enhancing interactions between Rydberg atoms is a key challenge in contemporary quantum technologies. Stronger interactions enable faster Rydberg gates in digital processors and larger entangled states in analog simulation. Achieving the same interaction strength at lower principal quantum number addresses current constraints in available Rabi frequency and field sensitivity in large scale tweezer or cavity QED experiments. Here, we demonstrate a new technique using AC Stark shifts from a microwave drive to tune into a \foerst resonance, thereby modifying the interaction scaling with distance from $1/R^6$ to $1/R^3$. We validate enhanced Rydberg interactions (in strength and range) by probing cavity Rydberg polariton blockade at $n=44$ in $^{87}$Rb, improving from \gtwon in the Van-der-Waals regime to \gtwof in the dipolar regime on the \foerst resonance. Importantly, our technique allows minimal shifts of the original Rydberg state, suppressing detuning errors in gate protocols while maintaining quadratic insensitivity to DC electric fields. 
\end{abstract}

\maketitle

\section{Introduction}
\label{sec:intro}
Rydberg interactions are the cornerstone of modern neutral atom platforms for digital quantum computation and analog quantum simulation. 
Rydberg electrons provide extremely strong atom-atom interactions at micron-scale separations. The electrons in these highly excited states of two atoms interact with each other through a mutual, quantum-fluctuation induced dipole moment~\cite{saffman_quantum_2010,Saffman2016,Browaeys2020}, also known as the Van-der-Waals (VdW) interaction. Combined with their long lifetime, Rydberg atoms enable high fidelity two-qubit gates, and the creation of strongly entangled many-body states, making neutral atoms a competitive route to scalable quantum computation and simulation. 

\begin{figure*}[ht!] %figure* produces a double-column figure in a two-column document
    %\centering
    \includegraphics[width=173 mm]{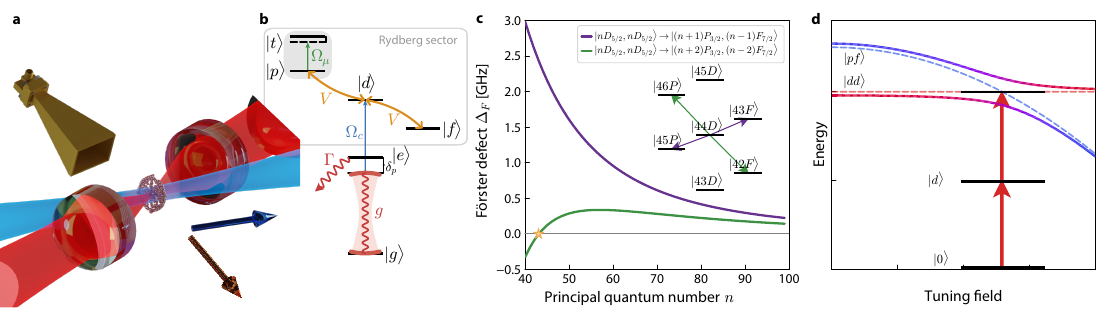}
    \caption{
        \textbf{Overview of experiment and \foerst Resonance mechanism}. \textbf{(a)} Schematic of the experimental setup: a cloud of ultracold $^{87}$Rb atoms is transported into a lens-focused cavity at $780$~nm (red) while a control field at $480$~nm (blue) excites to the Rydberg state. A microwave horn (gold) is used to inject RF fields that selectively shift different Rydberg states into a \foerst resonance. Arrows indicate quantization axis $\hat{z}$ (blue) and microwave propagation direction $\hat{k}_\mu$  (orange). \textbf{(b)} The atomic ground $\ket{g}$, excited $\ket{e}$, and Rydberg $\ket{d}$ state are coupled by a cavity mode with collective coupling $g$ and a classical control field $\Omega_c$ to realize cavity electromagnetically induced transparency (EIT). The Rydberg state $\ket{d}$ is dipole-dipole coupled to two other Rydberg states $\ket{p}$ and $\ket{f}$ with strength $V$, where $\ket{p}$ is AC Stark shifted by driving a tuning transition to a state $\ket{t}$ far off resonance with microwave Rabi frequency $\Omega_\mu$. \textbf{(c)} \foerst defect $\Delta_F$ as a function of principal quantum number $n$ for the two dominant channels shown in purple and green (see inset for $n=44$). At $n=43$ (gold star) a natural \foerst near resonance $\Delta_F \approx 8$~MHz occurs. \textbf{(d)} Blockade mechanism for tuned \foerst resonance: In the Van-der-Waals regime at zero tuning field, the defect between pair states $\ket{dd}$ (red) and $\ket{pf}$ (blue) is large and leads to only a small interaction shift (solid line) compared to the non-interacting case (dashed line). With increasing tuning strength and resulting $\ket{pf}$ shift, the interaction becomes dipolar and the two pair states show an avoided crossing at $\Delta_F=0$; the splitting is given by the dipolar interaction strength and the eigenstates are symmetric superpositions, indicated by the color. While the single-excited Rydberg state $\ket{d}$ is unperturbed, the second Rydberg excitation is now strongly blockaded from entering the system.
    }
	\label{fig:SetupFig}
\end{figure*}

In principle, operating with Rydberg atoms at high principal quantum number $n$ is desirable because the interaction strength increases as $n^{11}$ and lifetime as $n^7$. However, the large electron dipole moment enabling the interaction also makes Rydberg excitations very susceptible to environmental electric fields~\cite{Samutpraphoot2020nano}. While uniform background fields and gradients can be compensated using electrodes~\cite{clark_observation_2020,Panja2024} or passively shielded using Faraday cages, patch charges from nearby surfaces generate electrostatic potentials~\cite{Patrick2025,ocola_control_2024} that vary in time. The latter are difficult to compensate and consequently broaden the Rydberg state. This is a limiting factor already in current glass cell experiments, where surface charges need to be carefully controlled using UV light~\cite{Mamat2024}, and will become even more challenging for integration with photonic micro-structures~\cite{ocola_control_2024} or micro-cavities. Another challenge -- in particular for large atom array experiments -- is the increasing optical power required for Rydberg excitation with increasing $n$, due to the decreasing transition strength matrix element, scaling as $n^{-3/2}$.

Fundamentally, two Rydberg atoms in the same state have no direct dipole-dipole interaction~\footnote{In the absence of any polarizing field and for states of well-defined parity.} as the odd parity of the dipole operator makes the interaction vanish for the even product of identical states. Therefore, any interaction is second-order through an intermediate state, making typical Rydberg interactions of the Van-der-Waals type ($1/R^6$), rather than dipolar ($1/R^3$). Already early on, it was proposed that the dipolar nature of the interaction can be restored~\cite{Safinya1981,Ryabtsev2010,Nipper2012,ravets_coherent_2014} by ensuring that the intermediate pair state is at the same energy -- that is, its \foerst defect $\Delta_F$ is zero. This is called a \foerst resonance, and while it can occur naturally, for the overwhelming majority of chosen atomic states, it must be tuned through external fields. Previously, DC-electric fields have been employed to tune to the \foerst resonance, exploiting differences in the atomic polarizability $\alpha$ between different atomic states $l,m_l$~\cite{le_kien_dynamical_2013, bohlouli-zanjani_enhancement_2007, booth_reducing_2018, petrosyan_binding_2014, sevincli_microwave_2014, tretyakov_controlling_2014, kurdak_enhancement_2025} akin to Zeeman tuning to Feshbach resonances~\cite{chin2010feshbach}.
As the relative differences in polarizability $\alpha$ are small between nearby Rydberg levels, a large common-mode shift accompanies the tuning, which linearly amplifies even small fluctuations of the DC field, broadening the Rydberg level.
% Added one sentence here that mentions the previous appraoches. @Lukas: Maybe you can make it nicer and make sure that the sentence includes all appraoches (or at least most).

% Historically, DC-electric fields have been employed to tune to the \foerst resonance, exploiting differences in the atomic polarizability $\alpha$ between different states $l,m_l$, akin to Zeeman tuning to Feshbach resonances~\cite{chin2010feshbach}. Many experiments have demonstrated the use of DC-Stark shifts using static fields~\cite{le_kien_dynamical_2013, bohlouli-zanjani_enhancement_2007, booth_reducing_2018, petrosyan_binding_2014, sevincli_microwave_2014, tretyakov_controlling_2014, kurdak_enhancement_2025}, and it has been popular due to its simplicity. However, DC fields have drawbacks: the Stark effect is quadratic, so only defects of a particular sign can be tuned away. Furthermore, because the relative differences in polarizability $\alpha$ are small, a large common-mode shift accompanies the tuning, resulting in a shift of the single-atom excitation resonance. Perhaps most importantly, in the presence of a tuning field $E$, the quadratic Stark Hamiltonian $H = - \alpha |E|^2$  amplifies stray background fields and gradients linearly.

To circumvent analogous challenges in magnetic Feshbach resonances, optical Feshbach resonances were introduced, providing spatial control of interactions~\cite{clark_quantum_2015} and Feshbach-tuning of magnetic field insensitive atoms~\cite{nicholson_optical_2015}. While optical Feshbach resonances come at the cost of increased loss because of coupling to short lived excited states, \foerst resonances do not have this problem as other Rydberg states have similarly long lifetimes.

Here, we experimentally tune the interactions in a cavity Rydberg polariton experiment to a \foerst resonance using radio frequency (RF) fields, via the AC Stark effect. We characterize the increased interaction strength by directly measuring the RF-\foerst enhanced photon blockade of $^{87}$Rb at $n=44$, where the $g^{(2)}$ is enhanced from \gtwon to \gtwof. Our method provides several key benefits compared to previous~\cite{Safinya1981,Ryabtsev2010,Nipper2012,ravets_coherent_2014} approaches: (1) Strongly reduced sensitivity to DC electric fields; (2) bi-directional shifting independent of defect sign; (3) Fast switching of the interactions via electronic (RF) control.

% Not sure yet, where to put this:
% In quantum simulations experiments Rydberg interactions enable the use of \textit{super-atoms}. Working with ensembles of atoms yields high collective cooperativities while the large blockade radius restores the non-linearity needed for many-body states.

%  END OF THE INTRODUCTION

\section{Dipole-Dipole Interactions and \foerst Resonances}

Our experimental setup consists of a running-wave optical cavity (Finesse $\mathcal{F}=1200$) with a mode waist of $\sim$18~$\mu$m. We load a two-dimensional cloud (80$\times$80$\times$10~$\mu$m) yielding approximately $1.5 \times 10^3$ $^{87}$Rb atoms in the fundamental mode waist of the cavity and perform our experiments by measuring the out-coupled photons. The cavity (at $780$~nm) is resonant with the ground to excited state ($5$S$_{1/2}\leftrightarrow 5$P$_{3/2}$) transition, and an additional control field (at $480$~nm) excites electrons to Rydberg states realizing electromagnetically induced transparency (EIT). The eigenstates of the hybrid system are polaritons, superpositions of a collective atomic excitation to a Rydberg state (shared over many atoms) and a photon in the cavity. Our experiment uses long-lived dark polaritons as the quasiparticles, whose Rydberg state part interacts and ultimately blockades the cavity mode~\cite{georgakopoulos_theory_2018}. 
\begin{figure*}[ht!] %figure* produces a double-column figure in a two-column document
	\centering
    \includegraphics[width=173 mm]{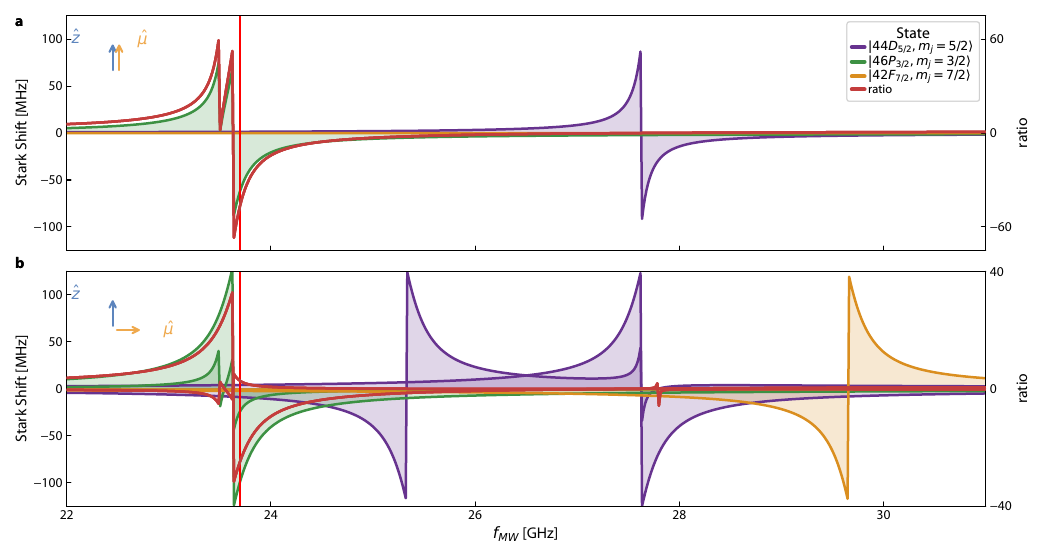}
	\caption{\textbf{AC Stark shift calculation of Rydberg states.} We explore the optimal way of tuning the pair state $\ket{pf}=\ket{(n+2)P_{3/2},(n-2)F_{7/2}}$ into resonance with our target state $\ket{dd}=\ket{n D_{5/2},n D_{5/2}}$ while minimizing the shift of the latter one. We compute the AC Stark shift for the three Rydberg states (color) over a wide range of microwave driving frequencies for $\pi$- \textbf{(a)} and $\sigma$- \textbf{(b)} polarization at a field strength of $E=0.1$~V/m, see \cref{SI:theory}. Driving with $\pi$ polarization is optimal as it maximizes the frequency separation of the $\ket{(n+2)P_{3/2}}$ from the target state and eliminates any vector shifts that would cause a transverse coupling between magnetic sublevels. The microwave driving frequency used in the experiment of $f_{MW}=23.7$~GHz is indicated by the vertical red line.}
	\label{fig:acstark}
\end{figure*} 
The interactions between the polaritons are entirely dictated by their Rydberg component: Rydberg atoms interact with each other through the electric dipole moment according to the dipole-dipole operator
\begin{equation}
    \hat{V} = \frac{1}{4 \pi \epsilon_0} \frac{\hat{\mu}_j \cdot \hat{\mu}_k - 3 (\hat{\mu}_j \cdot \vec{n}) (\hat{\mu}_k \cdot \vec{n})}{R^3_{j,k}}
\end{equation} where $R_{j,k}$ is the separation between atom $j$ and $k$, $\vec{n}$ is the unit vector along the interatomic axis and $\hat{\mu}_{j,k}$ is their respective dipole operator. Note that in our system, the atoms are confined to a 2D plane perpendicular to the quantization axis (see~\cref{sup:axialtrapping}).

The nature of this interaction depends heavily on the internal states of the two atoms under consideration. In most single-species tweezer or cavity QED experiments, only a single Rydberg state is optically excited to mediate quantum gates or engineer many-body Hamiltonians -- we will refer to this Rydberg state as the \textit{target} state.
For a pair of atoms in identical states there is no direct dipole-dipole interaction because of the odd parity of the dipole operator. These atoms can then only interact via a second order coupling through an intermediate, virtual pair state, leading to the familiar Van-der-Waals type interaction $1/R^{6}$. In the following, we consider only a single virtual pair state or \textit{channel} representing the dominant state out of all contributing pair states, to convey the relevant physics in a simple picture. In our EIT scheme (see \cref{fig:SetupFig}) the \textit{target} state is a $D$-state and the relevant pair state is the symmetric superposition of the $P$ and $F$ states, which we denote by $\ket{pf}$, a more detailed treatment is provided in \cref{si:interaction}. In general, this state can be any pair of states conserving total angular momentum; furthermore, this treatment is equally valid for any target and pair state conserving angular momentum~\cite{walkerConsequencesZeemanDegeneracy2008}.
%whether the target state is an $S$ or $P$ state. 

In the pair-state basis $\{\ket{dd}, \ket{pf} \}$, the Hamiltonian is given by:
\begin{equation}
    H = \begin{pmatrix} 0 & C_3/R^3 \\ C_3/R^3 & \Delta_F \end{pmatrix},
\end{equation}
where $\Delta_F = 2 E_d - E_f - E_p$ is the \foerst defect due to quantum defects for the different $l$ states and $C_3 = \bra{d}\hat{\mu}\ket{p} \bra{d}\hat{\mu}\ket{f}/4 \pi \epsilon_0$ quantifies the interaction strength.
\label{sec:acstark}
\begin{figure}[ht!] %figure* produces a double-column figure in a two-column document
    \centering
    \includegraphics[width=0.92\columnwidth]{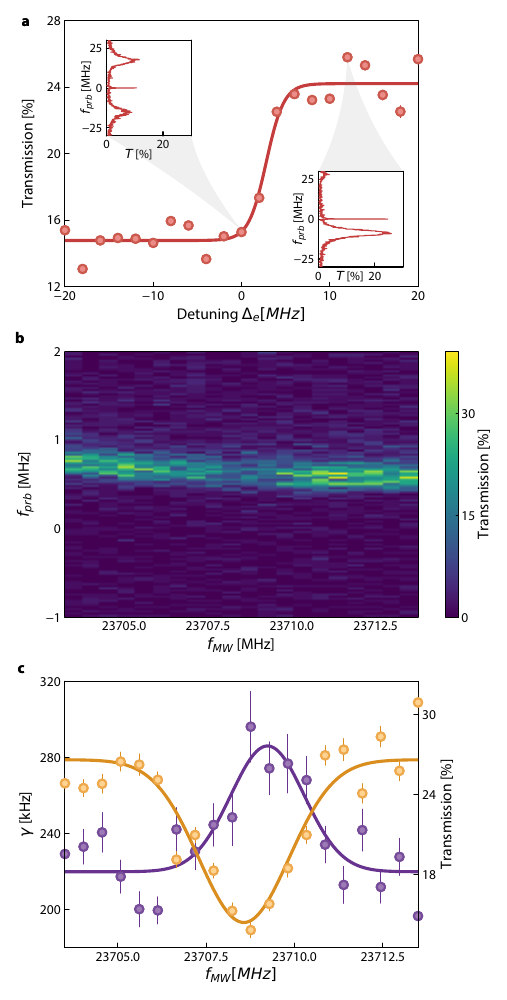}
    \caption{
        \textbf{Locating the \foerst Resonance: Mean field non-linearity}. The condition for the \foerst resonance $\Delta_F=0$ is narrow and needs careful tuning of the microwave frequency and power. To achieve this, we measure the mean-field interaction between Rydberg polaritons. \textbf{(a)} Depending on the intermediate state detuning $\Delta_e$ in our EIT scheme, the character of the interaction can be changed. On resonance ($\Delta_e=0$, left inset) the interaction potential is mostly imaginary leading to loss of additional polaritons, while at positive detuning the interactions are predominantly real and lead to a dispersive shift of the transmission peak and increased overall transmission (right inset). The interaction regimes are highlighted with a sigmoid function (solid line). \textbf{(b)} We find the exact resonance condition by measuring the EIT spectrum as a function of the microwave frequency, $f_{MW}$. \textbf{(c)}  Corresponding Lorentzian fits to the data in \textbf{(b)}  show a dip in transmission (orange) and a peak in line width (purple) at \foerst resonance.
        % We measure the exact resonance condition by observing a reduction in transmission \textbf{(b)} and an increase in dark polariton linewidth \textbf{(c)} as we vary the microwave drive frequency. 
        % \textbf{(c)} At very low probe powers $P_{prb}$ with less than one polariton per cavity mode $n<1$ we see a linear dependence of cavity transmitted power on the incident power, with a proportionality constant $T$; at higher  probe power $n>1$ the transmitted power saturates due to mean-field interactions between polaritons. The \foerst resonance increases both the interaction strength and range, so the non-linearity sets in at lower incident power.
    }
	\label{fig:nonlin}
\end{figure}
Depending on the spacing $R$ or size of the defect $\Delta_F$, the character of this Hamiltonian falls into two regimes:
(1) In the non-resonant interaction regime $\Delta_F\gg V(R)$, the eigenstates are largely unperturbed and the energy of the target pair state is shifted by
$-C_6/R^6$, leading to the typical Van-der-Waals interactions $1/R^{6}$. This dispersive shift is the standard case for Rydberg atoms in typical experiments and can be seen in \cref{fig:SetupFig} (d) in the absence of a tuning field.

(2) In the resonant interaction regime $\Delta_F \rightarrow 0$, the eigenstates hybridize, becoming superpositions of the target and virtual states $\frac{1}{\sqrt{2}}(\ket{dd} \pm \ket{pf})$. The interaction takes the form of a coherent dipolar exchange with a characteristic $1/R^{3}$ distance dependence (as given by the eigenenergies $\pm C_3/R^3$).
This leads to a different kind of Rydberg blockade shown in \cref{fig:SetupFig} (d) with two branches $\ket{dd} \pm \ket{pf}$ split symmetrically around the pair of target states $\ket{dd}$.

While natural \foerst resonances only occur in special, fine-tuned circumstances, or for very small atom separations, they can be highly desirable, as dipolar interactions decay much slower with increasing atomic distance $1/R^{3}$ compared to Van-der-Waals interactions $1/R^{6}$ and can therefore be much stronger at larger atom separations. In the following we will show how microwave (MW) driving can be used to intentionally tune the defect to zero and reach a \foerst resonance.

\section{RF Tuning $^{87}$R\lowercase{b} to a \foerst Resonance}
To transition from the Van-der-Waals to the dipolar regime, the \foerst defect $\Delta_F$ must be actively tuned to zero via external fields. Previously, DC electric fields have been used to tune the \foerst defect via the DC Stark effect. Since the energy spacing between Rydberg levels scales as $n^{-3}$ and dipole moments as $n^2$, the static polarizability, $\alpha_s$, exhibits a dramatic scaling of $\alpha_s \propto n^7$. While effective, DC Stark tuning has the unfortunate feature that all Rydberg states, including the target state, experience a large common-mode shift. DC electric fields thus cause a large Rydberg detuning that scales sensitively with the applied field strength. Hence, any fluctuating background field $\delta_E$ is linearly amplified by the DC field according to $(E + \delta_E)^2 \simeq E^2 + 2 E \delta_E$, significantly increasing susceptibility to stray electric fields.

To achieve state-selective energy shifts, we employ an AC Stark shift from a microwave field. By selecting a microwave frequency $f_{MW}$ near a Rydberg-Rydberg transition -- for instance, close to a transition of the $\ket{p}$ state but far detuned from the $\ket{d}$ and $\ket{f}$ states -- we can induce a large AC Stark shift on the $\ket{pf}$ manifold while leaving the $\ket{dd}$ target state essentially unperturbed. This differential shift allows us to tune $\Delta_F$ through zero, reaching the \foerst resonance dynamically (see~\cref{fig:acstark}). At the same time, small static field fluctuations are suppressed $(E + \delta_E)^2 = 0 + \mathcal{O}(\delta_E^2)$ quadratically when the DC fields are properly nulled $E=0$ (see \cref{SI:efield}). 

In our experiment the ``target'' state is $\ket{n~D_{5/2,m_J{=}5/2} }$ and the energetically closest suitable pair state $\ket{pf} \equiv \ket{(n+2) P_{3/2, m_J {=} 3/2} ~(n-2); F_{7/2, m_J{=}7/2}}$. While the defect is naturally small for $n=43$ with $\Delta_F=-8.4$~MHz, we focus on $n=44$ with a \foerst defect of $\Delta_F=65$~MHz instead to increase the contrast in interactions. We neglect any other channels (pair states), due to their significantly larger defects (see \cref{fig:SetupFig} (c)). 

To determine the optimal frequency and polarization of the microwave driving field, we compute the AC Stark shift of all participating states including their fine structure in \cref{fig:acstark} using the \textit{ARC}~\cite{sibalic_arc_2017} library (see \cref{SI:theory} for details) for $\pi$ and $\sigma$ polarization. In the experiment, we choose $\pi$ polarization for two reasons: (1) to avoid vector shifts that act as a virtual transverse Zeeman coupling between different magnetic sublevels;  and (2) to ensure that the microwaves are maximally off-resonant without driving other Rydberg-Rydberg transitions. Finally, we choose to mainly shift the $P$-state, that has fewer allowed transitions which are further apart in frequency than F-state transitions. The chosen microwave frequency of $f_{MW}=23.7$~GHz (indicated by the red vertical line in \cref{fig:acstark}) leads to a ratio of wanted (P,F-state) to unwanted (D-state) shift of $\sim50$. We point out that it is possible to adjust the polarization and frequency such that no D-state shift occurs (see SI). The existence of such a ``magic'' driving frequency makes this scheme suitable for Rydberg-blockade based gates that are otherwise sensitive to detuning errors.

% Therefore, the EIT feature only changes its resonance frequency very slightly.
% If a different target Rydberg state $|r\rangle$ can be chosen with more transitions close to the shifting transition, a zero-crossing of the target state AC Stark shift allows the interaction to be tuned without changing the $|r\rangle$ state energy at all \textbf{(bottom)}. The existence of a ``magic'' driving frequency makes this scheme suitable for Rydberg-blockade based gates that are otherwise sensitive to detuning errors.

\section{Probe non-linearity}
\label{sec:nonlin}

To experimentally tune into a \foerst resonance, precise calibration of microwave power, polarization, and frequency is necessary (see \cref{SI:microwave}). While 2-photon correlations provide an unambiguous signature of photon blockade, they are time-consuming to measure and hence a very inefficient method to find the \foerst resonance. Instead, we locate the resonance by using the mean-field nonlinearity of the transmitted probe light as a proxy for interactions. The general scheme of finding the \foerst resonance is as follows: we probe the system at a $780$~nm cavity photon flux that ensures that in steady-state multiple polaritons ($\sim 3$) are excited. When fine-tuning into resonance, the Rydberg blockade radius becomes larger (due to increasing interaction strength), and hence fewer polaritons can be simultaneously excited, resulting in a reduced cavity transmission.

% The probe non-linearity in our system occurs as the excitation volume available to polaritons is restricted in the axial direction by the small thickness of the cloud and in radial direction by the waist of the fundamental mode. When the Rydberg blockade radius is smaller than the waist, multiple polaritons can occupy the atomic sample, but for increasing interaction strength, fewer polariton can be excited, and the transmission plateaus as a function of the incident photon rate. At the low principal quantum numbers present, mesoscopic number of polaritons experience a mean-field energy shift that reduces but not fully saturates the transmitted intensity.

In order to increase the transmission contrast, we furthermore operate detuned from the intermediate $5$P$_{3/2}$-state. For our ladder EIT scheme, the exact interaction potential of two dark polaritons depends upon the intermediate state detuning $\Delta_e$~\cite{Bienias2014}. On resonance ($\Delta_e = 0$), the EIT condition is broken for the second polariton due to the Rydberg blockade from the Rydberg component of the first polariton. The second polariton is therefore no longer dark to the lossy intermediate state and can decay into free space, leading to a predominantly imaginary interaction potential between polaritons. Conversely, with a detuning off the intermediate state $\Delta_e \gg \Gamma$ the photon is mostly reflected off the cavity, as for a real valued repulsive interaction.

We operate with a small detuning $\Delta_e \approx 2 \Gamma$ where the interaction between polaritons leads both to a mean field shift of the two-photon resonance as well as an increased loss broadening the polariton linewidth. In \cref{fig:nonlin} (a) the increase in transmission for a positive P-state detuning $\Delta_p$ is seen to double the cavity transmission while maintaining large detuning from the bright polariton (see lower inset).% For maximum contrast we probe at an intermediate state detuning of $\Delta_e=12$~MHz.

Due to technical details specific to our setup (microwave power can only be tuned in steps; see \cref{SI:microwave}), we operate at a fixed microwave power that gives us the calculated $d$-state shift and then vary the microwave frequency $f_{MW}$ around the point indicated in \cref{fig:acstark} to tune exactly onto the resonance. The resulting dark polariton signal is shown in \cref{fig:nonlin} (b). By fitting Lorentzians to this data, we extract the transmission and linewidth as a function of the microwave frequency. As shown in \cref{fig:nonlin} (c) a reduction to half the transmission (green crosses) and simultaneous increase in linewidth (purple circles) for a narrow range in microwave drive frequency can be observed. From this we extract an experimental microwave frequency of $23708.6(1)$~MHz, showing a $\sim 10$~MHz deviation from the calculated parameters~\cite{sibalic_arc_2017}.

\section{Correlation function}
\label{sec:g2}

The true hallmark of strong single-photon-level non-linearity is the observation of photon blockade. In atom based quantum computation, the analogous Rydberg blockade enables two-atom gates; in materials made of light, photon blockade is the equivalent of Coulomb repulsion in real space electronic systems. We quantify the blockade by measuring correlations in the transmitted photons using single-photon-counting modules (SPCM) and tag the photon arrival times to compute coincidences.

\begin{figure}[ht!] %figure* produces a double-column figure in a two-column document
    %\centering
    \includegraphics[width=\columnwidth]{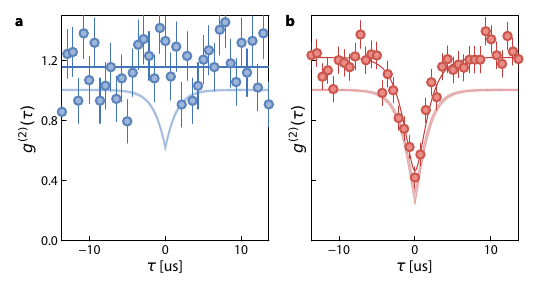}
    \caption{
        \textbf{Observation of photon blockade via \foerst resonance.} We measure the temporal correlation function $g^{(2)}(\tau)$ of photons transmitted through the cavity QED system. \textbf{(a)}  Away from the resonance, interactions in the $\ket{44D_{5/2}}$ state are insufficient to blockade the $LG_{0,0}$ mode, yielding \gtwon. \textbf{(b)} At the \foerst resonance, interactions are enhanced, resulting in a suppression of multi-photon events with \gtwof. The atomic cloud is confined to a quasi-2D geometry ($\sigma < R_b \lessapprox w_0$), ensuring that the blockade radius covers the transverse optical mode. Solid line is an analytical fit while shaded bands indicate \textit{ab initio} atomistic calculations (see \cref{SI:theory}) using experimentally measured cloud thickness $\sigma_z$. In \textbf{(a)} the microwave drive is still present but intentionally detuned in frequency to exclude effects of state admixture.}
	\label{fig:g2}
\end{figure}

The measured $\gtwo$ correlation function is shown in \cref{fig:g2} away from (a) and at the \foerst resonance (b). To rule out any microwave dressing or admixture effects, the microwave drive was present in both cases, but with its frequency shifted by $18$~MHz in (a) breaking the resonance. Plotted uncertainties are standard error in the mean. While no antibunching dip is visible for the detuned microwave drive (a), a clear dip is visible at \foerst resonance (b). Using a simple analytical model~\cite{clark_observation_2020} 
\begin{equation} \label{eq:g2model}
    G^{\text{mod}}(\tau) = G^{\text{class}} \left( 1 + (1 - G^{\text{min}}) (e^{-\gamma|\tau|} - 2e^{-\gamma|\tau|/2}) \right)
\end{equation} with $\tau$ the time delay between coincident photons and $\gamma$ the width of the blockade feature. We extract \gtwon in (a) without the \foerst resonance, and \gtwof in (b) when tuned to resonance. From $G^{\text{min}} = {\gamma^2}/{(\gamma^2 + U^2)}$ we can deduce an effective interaction strength of $U = 1.4(4)$~MHz. Note that the width of the correlation dip is dictated by the dark polariton linewidth $\gamma = \cos^2(\vartheta) \kappa +  \sin^2(\vartheta) \gamma_r$, which is comprised both of the cavity linewidth $\kappa$ and the collective Rydberg linewidth $\gamma_r$ according to the dark state rotation angle $\tan\vartheta = g/\Omega$. 

We compare the measured correlations with an atomistic, non-Hermitian perturbation theory (NHPT) (see \cref{SI:theory}) that estimates $\gtwo$ based on the theoretically calculated~\cite{sibalic_arc_2017} \cthree and \csix values. The results are shown in \cref{fig:g2} as shaded bands reflecting uncertainty in the exact cloud thickness, and give a $\gtwo$ of $g^{(2)}_{nhpt} = 0.61(1)$ and $g^{(2)}_{nhpt} = 0.28(4)$ for the two cases, respectively. The measured $\gtwo$ on \foerst resonance is compatible with the \textit{ab initio} calculations within the experimental errors, however the experimentally measured values of $g^{(2)}(\tau)$ at longer $\tau$ deviate from the expected value of 1 for a coherent state. These extraneous correlations can be caused by atom number fluctuations between- and within- experimental runs. We model these classical correlations at large time as $G^{\text{class}}$ in \cref{eq:g2model}. Note, however, these classical fluctuations can only increase the $g^{(2)}$ value.
The NHPT \textit{ab initio} theory overestimates the blockade in the absence of the resonance, as its truncated Hilbert space is not well suited to capture mean-field interactions.

From the theoretical model, we can furthermore deduce that the \foerst enhanced interactions at $n=44$ lead to a similar blockade strength to what we anticipate for Van-der-Waals interaction at $n=78$, while requiring less blue power (scaling $P \propto n^3$, indicates $5.6$ times less) for the same Rabi frequency, and reducing the DC polarizability (and thus field sensitivity) by a factor of 55.

\section{Summary \& Outlook}
\label{sec:outlook}
We have demonstrated a microwave tuned \foerst resonance, increasing interactions from $C_6 = 3.26 $~GHz $\mu$m$^6$ to $C_3 = 1.32 $~GHz $\mu$m$^3$ at principal quantum number $n=44$ in $^{87}$Rb. At the characteristic distance of the cavity mode waist $w_0=18$~$\mu$m, we estimate that the interaction energy increases from $U_6 = 100$~Hz to $U = 1.4(4)$~MHz, while measuring an enhanced photon blockade from \gtwon to \gtwof. 
The AC Stark drive only leads to a small residual shift of the optically excited, target $D$-state of $200$~kHz while the pair state was shifted by $62$~MHz, making this scheme applicable to higher principal quantum numbers.
Microwave \foerst tuning provides excellent state selectivity, far beyond what is achievable with DC tuning. Nonetheless, high interaction contrast is most readily achieved in the vicinity of a natural \foerst resonance, where only small Stark shifts (compared to the intrinsic level spacings) are required. Absent a nearby \foerst resonance, both the defects and level spacings scale as $n^{-3}$~\cite{walker_consequences_2008}, making the requisite strong RF drive less state selective and opening the Rydberg Hilbert space through multi-photon resonances, leading to dephasing.

Our scheme paves the way to use \foerst resonances as a means to further increase interactions at higher quantum numbers. In particular, in cavity QED experiments this will enable stronger blockade, especially cross-mode blockade in multi-mode experiments~\cite{clark_observation_2020} at reduced electric field sensitivity. Fast switching of interactions could enable Rydberg enhanced detection~\cite{gorniaczyk_enhancement_2016,xu_fast_2021,vaneecloo_intracavity_2022}. In multi-species experiments~\cite{beguin_direct_2013, walker_consequences_2008, ravets_coherent_2014, barredo_coherent_2015, ireland_interspecies_2024, du_imaging_2024, walker_consequences_2008-1} our technique enables a wider range of inter- and intra-species resonances used for gates between data and ancilla qubits or even shielding of one species from interactions through use of a \foerst-zero~\cite{walker_zeros_2005}. In neutral atom tweezer experiments, increased interactions could improve gate fidelities~\cite{tsai_benchmarking_2025} and reduce susceptibility to surface charges in hybrid platforms, especially those employing nanophotonic integration. Furthermore, the increased range of the dipolar interaction could enable connectivity beyond nearest-neighbor in tweezer arrays. This potentially allows for multi-qubit gates, such as Toffoli or nCNOT gates, drastically reducing required circuit depth~\cite{Pelegri2022,Jandura2025,Petrosyan2025}.

\section{Acknowledgments}
This work was supported by AFOSR MURI Grant FA9550-19-1-0399, and AFOSR Grant FA9550-181-0317. Y.C.F. acknowledges support from the Stanford Graduate Fellowship in Science and Engineering. We would like to thank Bryce Gadway for stimulating discussions.

% \section{Author Contributions}
% The experiments were designed by L.P. The apparatus was built by L.P. and B.L. All authors analyzed the data and contributed to the manuscript.

% $^*$ These authors contributed equally.

\section{Competing Interests}
J.S. is a consultant for and on the advisory board of Atom Computing.

\clearpage

\putbib
\end{bibunit}

\subsection{Data Availability}
The experimental data presented in this manuscript are available from the corresponding author upon request, due to the proprietary file formats employed in the data collection process.
\subsection{Code Availability}
The source code for simulations throughout is available from the corresponding author upon request. 
\subsection{Additional Information}
Correspondence and requests for materials should be addressed to L.P. (lukasp@stanford.edu). Supplementary information is available for this paper.

\clearpage
\newpage
% \mbox{~}
% \clearpage
% \newpage

%\begin{bibunit}
% \begin{refsection}

\begin{bibunit}
\onecolumngrid
\newpage
\section*{Supplementary Information}
\appendix
\renewcommand{\appendixname}{Supplement}
\renewcommand{\theequation}{S\arabic{equation}}
\renewcommand{\thefigure}{S\arabic{figure}}
\renewcommand{\figurename}{Supplemental Information Fig.}
\renewcommand{\tablename}{Table}
\setcounter{figure}{0}
\setcounter{table}{0}
\numberwithin{equation}{section}

\section{Experiment details}
\label{SI:setup}
In this section, we present the experimental details about the setup, EIT scheme and probe sequence.

Our hybrid experimental platform consists of a multi-mode optical cavity and a cold atom source.
\begin{figure}[ht] %figure* produces a double-column figure in a two-column document
	\centering
    \includegraphics[width=0.6\textwidth]{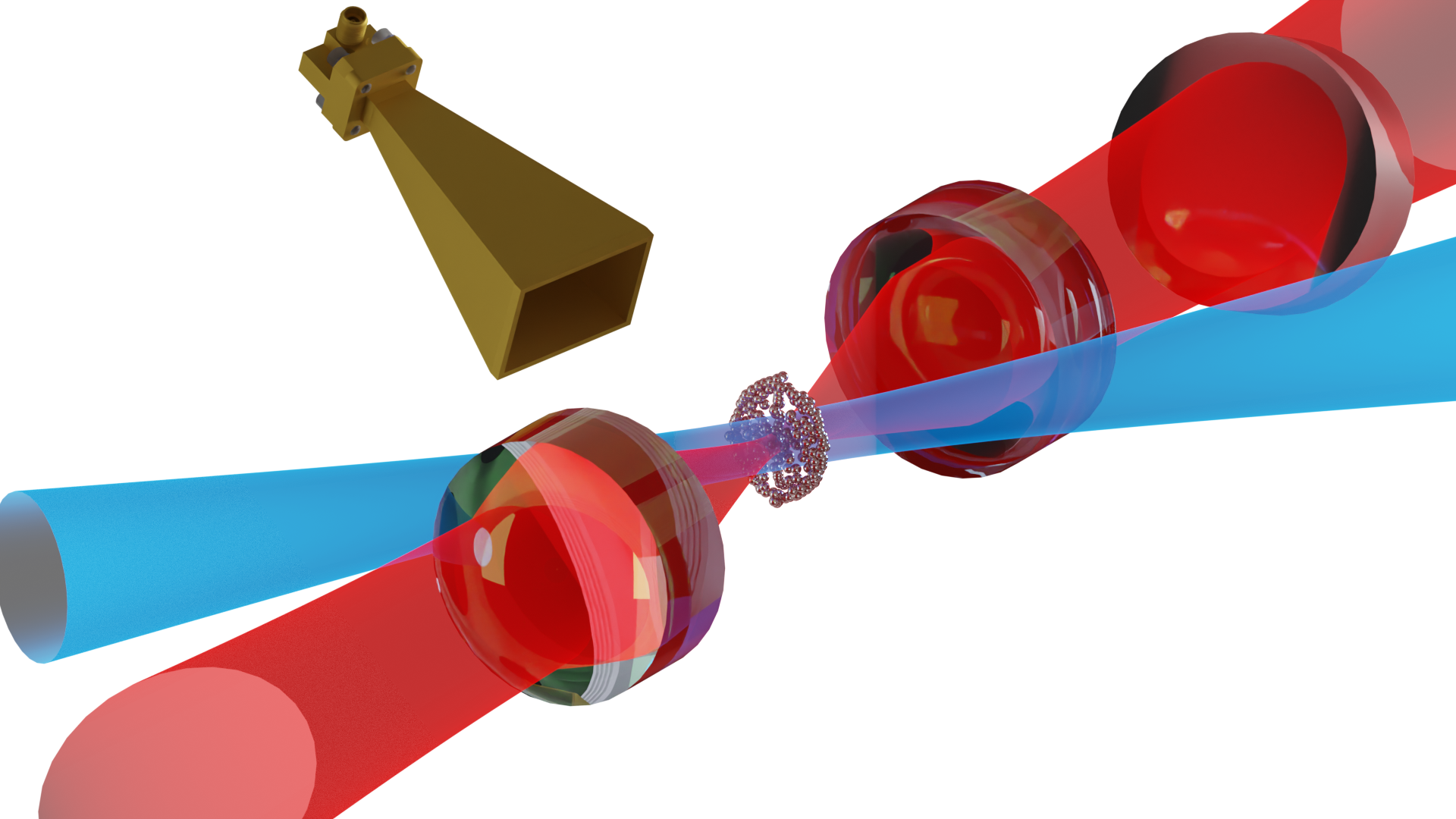}
	\caption{\textbf{Experiment Setup} Running wave cavity formed by four flat mirrors (only one shown) and two intra-cavity lenses at $780$~nm (red) with crossed buildup cavity (not shown) for the $480$~nm Rydberg excitation (blue). The atomic cloud (stylized atoms) is held in an elliptical lattice (light red) to confine them to a \textit{quasi} two-dimensional system. Microwaves for tuning the pair state are sent in via a horn transversely to the quantization axis.}
	\label{fig:setup}
\end{figure} 
The optical cavity in our experiment has a Finesse of $\mathcal{F} \approx 1200$ and a fundamental mode waist of $w_0 = 18$~$\mu$m. Although the cavity is specially designed with many higher-order transverse modes, in this work we only make use of the fundamental $LG_{0,0}$ mode and operate at a transverse mode splitting of $70$~MHz to freeze out any transverse degrees of freedom.
We prepare an ultracold ensemble of $^{87}$Rb atoms by collecting atoms from a background pressure in a 3D magneto-optical trap (MOT) and subsequently cool them using polarization-gradient cooling (PGC) to a temperature of $\sim 5$~$\mu$K. Optical molasses also facilitates loading into a vertical lattice that transports atoms into the waist of the optical cavity through an optical conveyor belt technique. At the cavity location, the atoms are further cooled in a 3D lattice using degenerate Raman sideband cooling (dRSC) to a temperature of $\sim 500$~nK and optically pumped to the $\ket{g} \equiv \ket{5S_{1/2,F{=}2, m_F {=} 2}}$ stretched state.
During probing, the atoms are held in the same vertical lattice, however at much reduced lattice depth of a few $\mu$K to minimize inhomogeneous broadening. The density of the atomic cloud in the center is on the order of $5 \times 10^{11}$~cm$^{-3}$, resulting in a collective coupling strength of up to $g = 50$~MHz. 

\subsection{Axial atom trapping}
\label{sup:axialtrapping}
\begin{figure}[ht] %figure* produces a double-column figure in a two-column document
	\centering
    \includegraphics[width=0.7\textwidth]{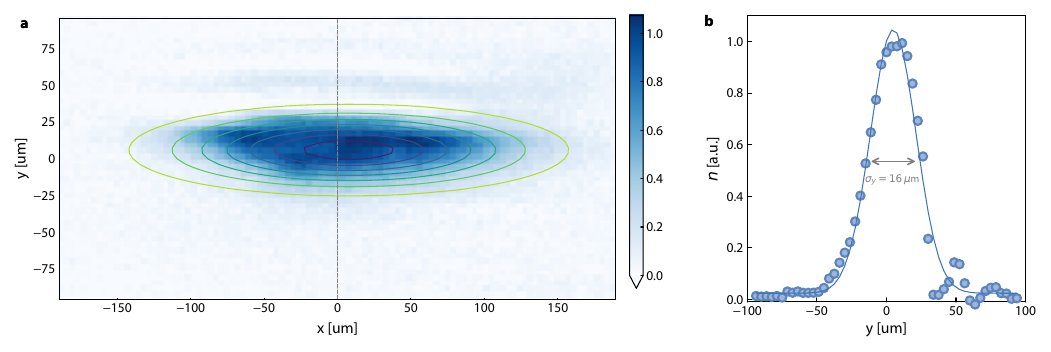}
	\caption{\textbf{Absorption image of the atom cloud} in the compressed, elliptical lattice. An integrated column density (a) and cut (b) reveal a size of the atomic cloud along the cavity direction (indicated in red) of $\sigma_z = 16(1)$~$\mu$m. The actual size of the cloud might be less as the imaging part has a small numerical aperture and is clipping.}
	\label{fig:si_abs}
\end{figure}

To achieve full Rydberg blockade of the ensemble, the blockade radius $R_b$ must be larger than the extent of the participating atoms. While the cavity waist limits the radial size of the atomic excitation, the longitudinal extent of the sample along the cavity axis still allows for multiple Rydberg excitations in the system and prevents full blockade. We therefore transfer the atoms from the symmetric transport lattice into a special, \textit{elliptical}, lattice with astigmatic waists $\omega_{x} \approx 200$~$\mu$m and $\omega_{y} \approx 20$~$\mu$m making the sample thin in the longitudinal direction and effectively 2D $\omega_{y} < w_{0}$. This lattice is generated by the same laser, but the beam is made highly astigmatic using a cylindrical telescope before focusing. The two lattices have opposite linear polarization and are detuned in frequency by $160$~MHz to prevent mutual interference.
After the atoms are transferred to the elliptical lattice with an adiabatic ramp, a second round of dRSC is performed to dissipate the increased kinetic energy from compression.
The resulting atomic sample is imaged in absorption in \cref{fig:si_abs} and the fitted waist of the atomic cloud in the thin y-direction is $16(1)$~$\mu$m. The actual size of the atomic cloud might be even smaller as the imaging is limited by the numerical aperture of the imaging path. 

\section{Microwave Calibration}
\label{SI:microwave}
We want to shift only one of the Rydberg states constituting the pair state into resonance without providing coupling to other magnetic sublevels. We therefore need the polarization of the microwave field to be as purely $\pi$-polarized as possible to not generate any vector shifts. 
While more sophisticated methods for 3D microwave polarization control exist~\cite{kurdak_enhancement_2025}, delivering fields to in-vacuum electrodes becomes increasingly challenging due to the limited availability of coaxial cables and vacuum feedthroughs at mm-wave frequencies. We therefore choose a much simpler approach relying on the intrinsically linear polarization of the waveguide coupler and horn used in our experiments and the Gaussian-beam-like propagation of microwaves at these high frequencies. Microwaves in the mm band behave much like \textit{optical} Gaussian beams and a microwave horn generates an approximately Gaussian beam with a waist close to its aperture. We can then use optics such as dielectric lenses to manipulate the microwaves and focus them onto the atomic sample, which simultaneously maximizes Rabi frequency and minimizes reflections from the metal cavity structure and chamber that could scramble the polarization at the atoms.
\begin{figure}[ht] %figure* produces a double-column figure in a two-column document
	\centering
    \includegraphics[width=0.6\textwidth]{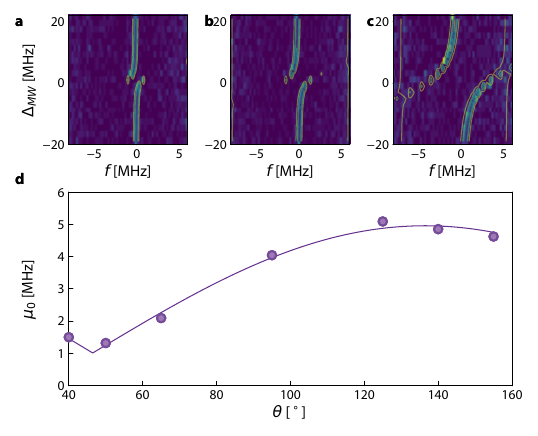}
	\caption{\textbf{Microwave polarization calibration} (a) - (c) Autler-Townes splitting of the EIT peak due to applied microwaves for three different horn orientations. Scanning the detuning $\Delta_{MW}$ of the applied microwave frequency reveals splitting of the dark polariton peak with increasing microwave drive power, analytical fit overlaid as contours. (d) Extracted microwave Rabi frequency $\mu_0$ as a function of horn rotation $\theta$. }
	\label{fig:microwave}
\end{figure} 
The microwaves at $\approx 24$~GHz are generated using a Vaunix Labbrick LMS-203 synthesizer, frequency doubled using a MiniCircuits ZXF90-3-443-K+ doubler, amplified by a Qorvo TGA4533-SM amplifier up to a power of 30 dBm and sent into the experiment using a coaxial waveguide adapter and horn.
As access into the experiment vacuum chamber is limited due to a relatively small viewport and the cavity structure itself, estimation of the electric field and polarization is hard as many internal reflections in the chamber could change field strength through interference and scramble the polarization. We therefore calibrate both using microwave transitions from the $\ket{44D_{5/2,m_J{=}5/2}}$ state. Constrained by access into the cavity structure that only has a large opening in the vertical direction, the microwave propagation direction $\hat{{k}}_{MW} \perp \hat{z}$ is perpendicular to the quantization axis set by a Zeeman field. For linearly polarized microwaves that the horn emits, they can therefore be $\pi$ or lin$\perp$lin polarized in the frame of the atoms.
The resulting atomic transitions have Rabi frequency $\mu = \{ \mu_-, \mu_\pi, \mu_+ \} = \mu_0  \{ \frac{\sin\theta}{\sqrt{2}}, \cos \theta, \frac{\sin\theta}{\sqrt{2}}  \}$ depending on the horn rotation angle $\theta$. First, we set the frequency to drive from the $\ket{44D_{5/2,m_J{=}5/2}}$ state to the $\ket{43 P_{{3}/{2},m_J{=}{3}/{2}}}$ state that can only be driven if the microwave field has an undesired $\sigma^-$ component due to dipole selection rules.
We then minimize this component through rotation and pointing of the horn and fit the resulting Autler-Townes splitting of the EIT peak as a function of angle in \cref{fig:microwave}. The top three panels (a)-(c) show the Autler-Townes splitting of the EIT peak as a function of microwave drive frequency $f_{MW}$ for three different angles of the microwave horn. We fit an analytical function derived from linear NHPT (see \cref{SI:theory}) using an effective Hamiltonian 
\begin{equation}
\hat{H}_{eff} = \begin{pmatrix}
\delta_c & g & 0 & 0 \\
g^* & \delta_p & \Omega& 0\\
0 & \Omega^* & \delta_r & \mu\\
0 & 0 & \mu^*  & \Delta
\end{pmatrix}
\end{equation}
where $\delta_j = \Delta_j - \delta_{probe} + i\frac{\Gamma_j}{2}$ is the complex detuning for $j \in \{c, p, r\}$ cavity, atomic $5$P- and Rydberg state including real detunings $\Delta$ and lifetimes $\Gamma$, capturing all EIT and linewidth effects and yielding an effective microwave Rabi frequency $\mu$.
From the fit of $\mu_-$ in \cref{fig:microwave} we can infer the microwave polarization infidelity given the non-vanishing value of $1$~MHz at the optimal setting of $47^{\circ}$ leading to a contrast ratio of $10:1$.

\section{Theoretical model}
\label{SI:theory}
\subsection{Stark shift numerical calculation}

%static vs dynamic? Where to put this?
The static electric polarizability, $\alpha$, of an atom in a state $|r\rangle$ is a measure of the second-order energy shift, $\Delta E^{(2)}$, due to the presence of a constant electric field, $\mathcal{E} = \mathcal{E}_z \hat{z}$. The energy shift is given by:
$$
\Delta E^{(2)} = -\frac{1}{2} \alpha_{s} \mathcal{E}_z^2
$$
The polarizability tensor component $\alpha_{s}$ for the state $|r\rangle$ is then given by a sum over all states:
\begin{equation}
\alpha_s = 2 e^2\displaystyle\sum_{n'\ell'j'\neq n\ell j} \frac{\vert \langle n,\ell,j,m_j\vert r_0 \vert n',\ell',j',m_j' \rangle \vert^2}{E_{n'\ell' j'}-E_{n\ell j}}.
\end{equation}
where $E_{n\ell j}$ are the unperturbed energies.
The energies according to the Rydberg formula scale as $E_n \propto n^{-2}$ so their spacings scale as $n^{-3}$ and since the dipole matrix elements scale as $n^{2}$ the scalar polarizability scales as $n^{7}$.
We compute the static and dynamic polarizabilities using the ARC~\cite{sibalic_arc_2017} library using the Shirley method and include a window of $n\pm 5$
states around the state of interest with $l_{max}=10$ for the diagonalisation.

\subsection{Interactions}
\label{si:interaction}
% %old section
The intrinsic interaction strength between two Rydberg atoms is determined by the dipole coefficient $C_3$ and the \foerst defect $\Delta$. Two Rydberg states with identical quantum numbers experience no direct dipole-dipole ($1/R^3$) interaction because of the odd parity of the dipole operator. The interaction is therefore of a Van-der-Waals type ($1/R^6$) and can be understood as a second order interaction via a virtual pair state. The initial pair state (here, of identical states) couples to a pair of atoms that conserve total angular momentum, via the dipole operator. The strength of the second order VdW interaction is given by $C_6=C_3^2/\Delta$.

We model the interaction between a target Rydberg pair state, $|rr\rangle$, and a coupled pair state, $|r'r''\rangle$, using a simple effective Hamiltonian. Here we consider the target state to be a D state pair, $|rr\rangle=|dd\rangle$, coupled to a P-F state pair, $|r'r''\rangle=|pf\rangle$. The energy difference between these unperturbed pair states is $\Delta = E_{r'}+E_{r''}-2E_{r}$.
While both permutations must be accounted for, only the symmetric state $\ket{\tilde{pf}} = (\ket{pf} + \ket{fp})/\sqrt{2}$ couples to $|dd\rangle$ via the dipolar interaction, leading to the model Hamiltonian:
\begin{equation}
\mathcal{H} = \begin{pmatrix}
0 &V(R) \\
V(R)&\Delta
\end{pmatrix},
\end{equation}
leading to $$V_\pm(R) = E_{\text{avg}} \pm \sqrt{\left(\frac{\Delta}{2}\right)^2 + \left(\frac{C_3}{R^3}\right)^2}$$ %eigenvalues $ \lambda_\pm = \frac{\Delta\pm\sqrt{\Delta^2+4V(R)^2}}{2}$
where $V(R)$ is the dipole-dipole interaction. The behavior of the eigenvalues falls into two distinct regimes:
\begin{equation}
V_\pm(R)\approx
\begin{cases}
    -\dfrac{C_6}{R^6} & \text{for } |\Delta| \gg |V(R)| \quad \text{(Van-der-Waals regime)} \\
    \\
    \pm \dfrac{C_3}{R^3} & \text{for } \Delta \to 0 \quad \text{(Resonant regime)}
\end{cases}
\end{equation}
In the far-detuned (Van-der-Waals) regime, the energy shift is described by the $C_6$ coefficient, calculated by summing over the contributions from all coupled pair states:
\begin{equation}
C_6 = \displaystyle\sum_{r'r''} \frac{\left| \langle r'r''\vert V(R) \vert rr\rangle\right|^2 R^6}{\Delta_{r',r''}}
\end{equation}
In the case of resonance with a particular pair state ($\Delta_{r',r''}=0$), the eigenstates are split symmetrically by the resonant dipole-dipole interaction, which scales as $1/R^3$.

\subsection{Correlation function model}
In our continuum system, calculation of the blockade is more complicated than in discrete tweezer systems due to the range of atomic separations present within the cavity mode. We therefore perform an atomistic calculation of $N=400$ atoms but truncate the Hilbert space to two excitations. 

The interacting particles in our hybrid platform are really Cavity-Rydberg dark polariton, a quasi-particle of cavity photon and Rydberg excitation of the atomic ensemble. These polaritons are collective excitations of all atoms and they can interact through the large Rydberg component. When the photonic part of the polariton leaks out of the cavity, it retains the same correlations and blockade behavior as the underlying polariton.
Predicting blockade in this system is much harder than e.g. tweezer experiments, where Rydberg atoms are located at a well defined distance $R$. Instead, the Rydberg excitations in cavity QED are \textit{collective} in nature which means they are delocalized over the whole atomic sample leading to a distribution over a continuum in $R$ for every possible pair of Rydberg excited atoms. Therefore there is not a single well defined interaction energy that could be used to calculate the blockade, instead many atoms have large separations up to several cavity waists with small interactions whereas two atoms located right next to each other in the continuous sample have seemingly diverging interaction energies.
We therefore need to perform an effective calculation in order to predict the photon blockade strength. While effective theories for Rydberg polariton exist~\cite{gullans_effective_2016} in 1D, no self-consistent theory in 2 dimensions exists so far albeit attempts have been made~\cite{georgakopoulos_theory_2018}.

\begin{figure}[ht] %figure* produces a double-column figure in a two-column document
	\centering
    \includegraphics[width=\textwidth]{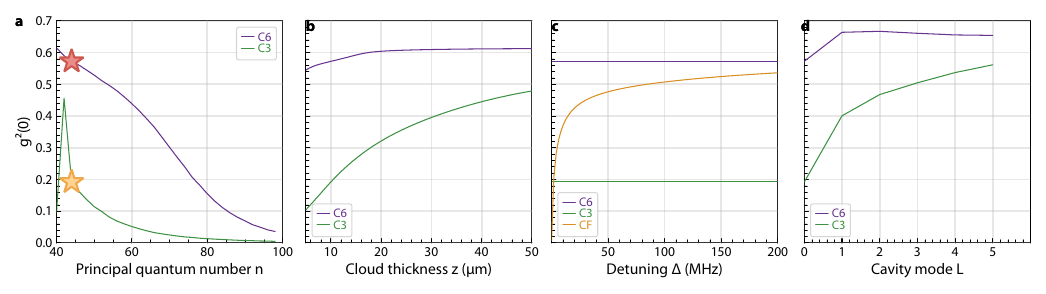}
	\caption{\textbf{Scaling of photon blockade $\gtwo$} with (a) principal quantum number $n$, (b) cloud thickness along the cavity axis $\sigma_z$, (c) \foerst defect and (d) cavity LG mode index $L$ for $n=44$. Theoretical values for the $\paird$ state used in this work are indicated as a star for untuned (red) and tuned (orange) \foerst resonance.}
	\label{fig:scaling}
\end{figure}

In \cref{fig:scaling}, the effect of photon blockade strength is calculated as a function of all relevant parameters. Specifically, the scaling for higher principal quantum numbers for both VdW \csix and dipolar \cthree regime is calculated in (a) for the current atom and laser parameters. While the state $n=44$ chosen in this work (stars) provides a high contrast between un-tuned and tuned interactions, the photons emitted from the cavity are not fully anti-bunched $\gtwo>0$. However, full blockade should already be reached at $n=78$ when employing the \foerst resonance that would otherwise require $n\approx110$ for a comparable blockade. This lower quantum number reduces the electric field sensitivity by a factor of $10$ and the required optical power for Rydberg excitation by $2.6$.
The second most important experimental parameter is the cloud thickness along the cavity direction $\sigma_z$ shown in \cref{fig:scaling} (b). 
The crossover behavior when tuning the \foerst defect from VdW (large $\Delta \gg 1$) to the dipolar ($\Delta \rightarrow 0$) regime on resonance is shown in \cref{fig:scaling} (c). 
We can eliminate the additional $\ket{\tilde{pf}}$ state that arises in the dipolar case analytically in frequency space and use the same NHPT to analyze the crossover.

\begin{figure}[ht] %figure* produces a double-column figure in a two-column document
	\centering
    \includegraphics[width=0.8\textwidth]{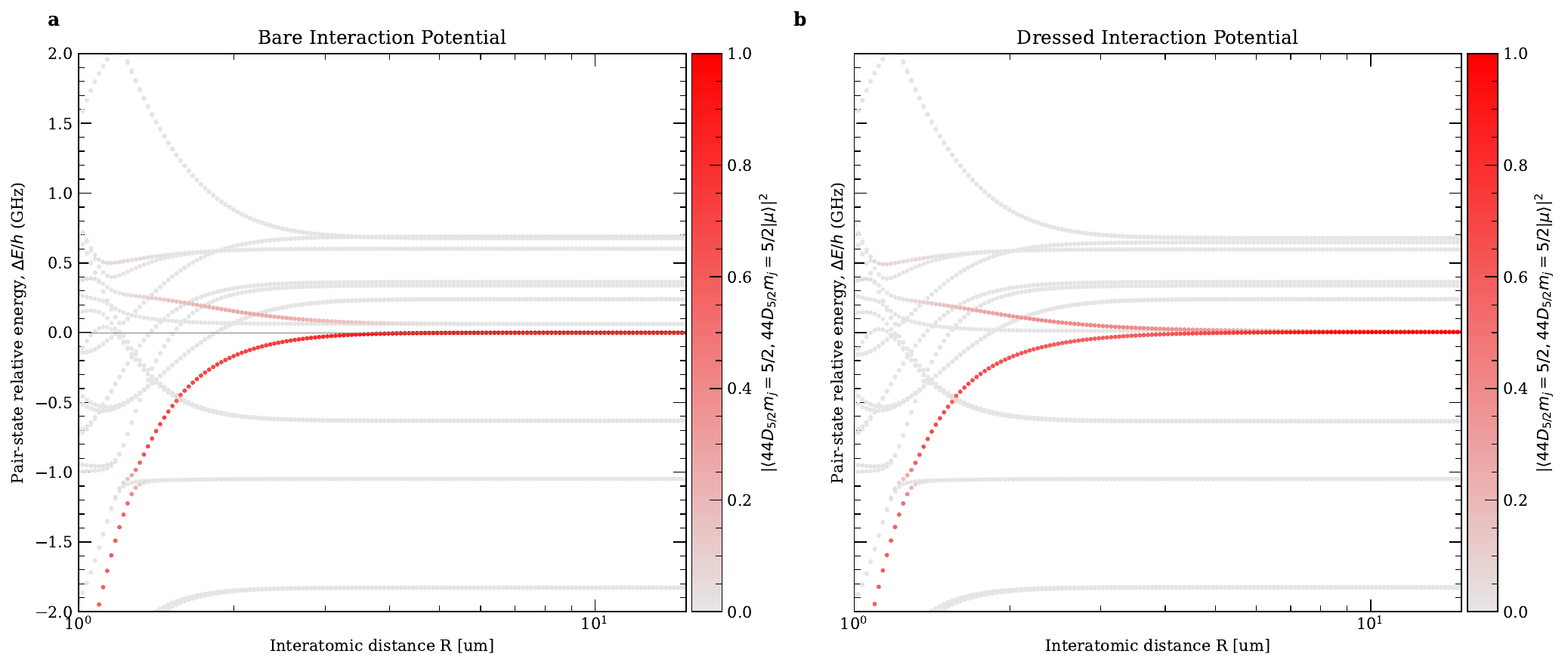}
	\caption{\textbf{Interaction potentials} for $\ket{44 D_{5/2,5/2}}$ (a) without and (b) MW tuning applied. The initial Van-der-Waals potential with only a small residual coupling to the upper branch at short distances, gets transformed into a dipolar potential with upper and lower branch split equally. In (b) AC Stark shifts on all states in the basis from \foerst drive are included to rule out accidental cancellations.}
	\label{fig:comparison}
\end{figure} 

\section{Interaction scaling}
\label{SI:scaling}

We compute Rydberg pair potentials for Rb atoms in state $|nL, j, m\rangle$ using the Pairinteraction~\cite{weber_tutorial_2017} library. The two-atom Hamiltonian is diagonalized at varying interatomic distances $R$ to obtain eigen-energies relative to the non-interacting pair energy. The energy basis is restricted to $E_{\text{pair}} \pm \epsilon_{\text{range}}$ where $\epsilon_{\text{range}} = 30(68/n)^2$~GHz scales with principal quantum number.
The pair potentials exhibit two distinct regimes characterized by different distance dependencies:
At large distances ($R > R_{\text{vdW}}$), non-resonant dipole-dipole interactions yield an \textit{effective attractive} potential
$$V_{\text{vdW}}(R) = -\frac{C_6}{R^6}$$
where $C_6 \equiv C_3^2/|\Delta|$ is the effective Van-der-Waals coefficient, with $\Delta$ the detuning from dipole-dipole resonance.
The Van-der-Waals radius $R_{\text{vdW}}$ is a characteristic length scale that marks the crossover from the $C_6$ regime (large $R$) to the $C_3$ regime (small $R$). It is determined by fitting the attractive eigenstate branch with highest overlap to the \textit{LeRoy-Bernstein model}
$$V(R) = V_\infty + A \frac{1 - \sqrt{1 + (R_{\text{vdW}}/R)^6}}{1 - \sqrt{1 + R_{\text{vdW}}^6}}$$
where $V_\infty$ is the asymptotic energy, $A$ is the well depth scale, and $R_{\text{vdW}}$ characterizes the position of rapid energy variation. This semi-empirical form analytically connects the long-range $-C_6/R^6$ behavior to the short-range avoided crossing. The crossover radius $R_{\text{vdW}}$ naturally emerges at the point of maximal energy gradient.

At short distances ($R < R_{\text{vdW}}$), near-resonant dipole-dipole coupling dominates, producing avoided crossings with characteristic $R^{-3}$ scaling where $C_3$ quantifies the dipole-dipole interaction strength.
We simultaneously fit one or both eigenvalue branches using the coupled two-state model
$$V_\pm(R) = E_{\text{avg}} \pm \sqrt{\left(\frac{\Delta}{2}\right)^2 + \left(\frac{C_3}{R^3}\right)^2}$$

This yields $C_3$ and $\Delta$ (detuning) directly from nonlinear least squares fitting. The effective coefficient $C_6 = C_3^2/|\Delta|$ follows from perturbation theory. For near-resonant (\foerst) cases, both branches are tracked and fitted simultaneously; for off-resonant cases, a single branch suffices. States with overlap $> 0.4$ with the initial pair state are preferentially included.

\section{Multimode cross blockade}
\label{SI:longrange}
The current experimental apparatus is specifically designed for studying many-body states under an artificial magnetic field in the lowest Landau level (LLL)~\cite{clark_observation_2020, clark_interacting_2019}. The different $LG_l^p$ modes form the orbitals that particles in the LLL ($p=0$) can occupy. To realize states of the fractional quantum Hall effect, strong interactions between the orbitals are required, which is the role of the Rydberg interactions studied in this work. To be able to build large correlated states, the interactions between all modes need to be strong to enable scattering between all the different modes. 
\begin{figure}[ht] %figure* produces a double-column figure in a two-column document
	\centering
    \includegraphics[width=0.6\textwidth]{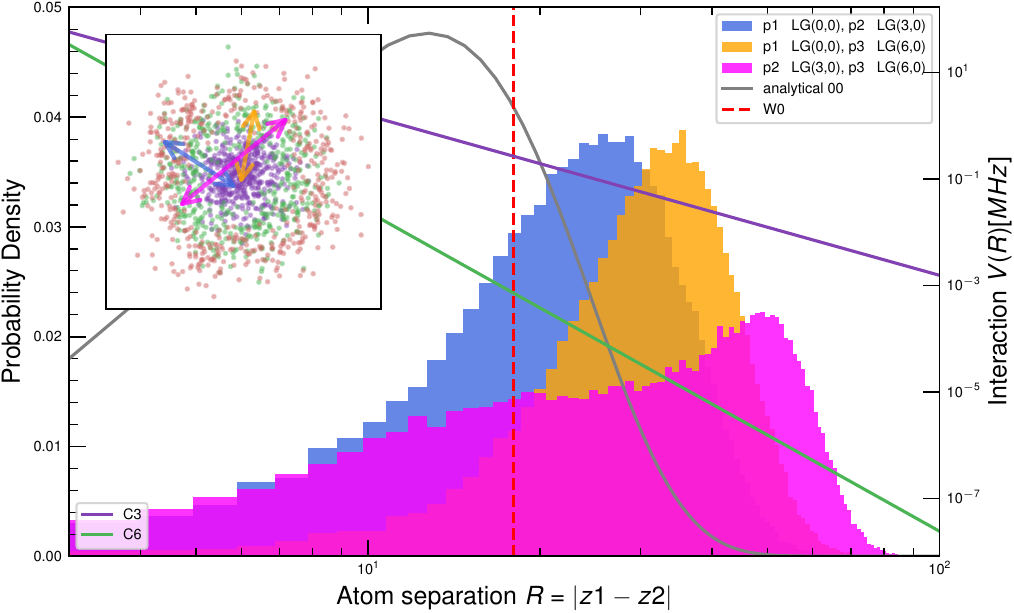}
	\caption{\textbf{Long range interactions should dramatically improve cross-blockade} Characteristic distances for all atomic pair distances for different modes (histograms) reveal that interactions at the range of the cavity waist dominate. Dipolar interactions (purple) can lead to a substantial enhancement over Van-der-Waals interactions (green) even though the latter dominate at short distances.}
	\label{fig:longrange}
\end{figure} 
As higher order $l$ modes sample a larger extent of the atomic sample, the expected distances between the atoms increase for higher $l$, where the exact distribution can only be found in closed form for the fundamental Gaussian mode $l=0$ and the higher order and cross-mode distributions can only be found numerically by Monte-Carlo sampling as shown in \cref{fig:longrange}.
The long range character of the dipolar $C_3/R^3$ interaction can lead to dramatically enhanced interactions by two to three orders of magnitude at the expected distances for higher order modes that are on the order of the cavity waist.

\section{Electric Field Cancellation}
\label{SI:efield}
Rydberg atoms, especially at high principal quantum numbers $n>60$ are extremely sensitive to static electric fields due to their large dipole moment. Even when the closest surface is $>1$cm away from the atom location, adsorbates, patch charges or electric potentials from piezos on the cavity structure can cause substantial background fields. If not controlled, these can lead to DC Stark shifts on the order of several MHz and as the exact charge distribution can change over time due to mobile charges/desorption, causing time-varying \textit{random} shifts. In the cavity used in this work, which relies on intra-cavity lenses for aberration control, large dielectric surfaces are close to the atoms, exacerbating this problem. While the exact mechanism for patch charge formation on dielectric vs metallic surfaces is not entirely understood, a common solution is the application of UV light which is theorized to desorb adsorbed Rubidium atoms from surfaces in glass cell experiments~\cite{Mamat2024,Patrick2025}. Due to limited optical access we shine a strong, collimated UV LED source at $365$~nm and $1$~W of power through one of the MOT view ports into the cavity structure illuminating all surfaces. When the UV light is applied during the \textit{steady state} between experiment runs it has no effect on MOT loading and we turn it off during the probe phase of every run. Active UV elimination stabilizes the background electric field after about $30$~s.
We cancel the remaining background field that is created in part by our cavity locking piezos as well as remaining charges using 9 in-vacuum electrodes. This is sufficient to cancel all absolute field components $E_i$ as well as the gradients $\frac{\partial E_i}{\partial x_j}$ at the location of the atoms as Maxwell's equations in free space enforce four additional constraints (divergence and curl free). Controlling all 9 electrodes individually would lead to a hard to optimize parameter space, also because their placement is not symmetric but governed by available space in the structure. We therefore perform a finite-element simulation of the cavity structure, lenses and electrodes in \textit{Ansys Maxwell} to extract a response matrix $\begin{pmatrix} E_i \\ \partial_j E_i \end{pmatrix} = M \begin{pmatrix} \vdots \\ V_k \\ \vdots \end{pmatrix}$
capturing the fields $E_i$ and gradients $\partial_j E_i $ by a voltage on electrode $V_k$.
This matrix can then be inverted to directly control the fields and gradients. A magnetic Zeeman field is applied to the atoms to break the hyperfine degeneracy and protect the EIT from slight imperfections in the polarization, which breaks the spatial symmetry and defines a quantization axis. The AC Stark shifts interact with \textit{simultaneous} Zeeman shifts, therefore the best decomposition of electric fields is the one parallel $E_{\parallel}$ and two perpendicular $E_{\perp},E_z$ axes.

% \begin{figure*}[t] %figure* produces a double-column figure in a two-column document
% 	\centering
%  	\includegraphics[width=0.8\textwidth]{figS2.png}
% 	\caption{\textbf{Wideband PDH locking} In order to improve the robustness }
% 	\label{fig:efield}
% \end{figure*} 
We optimize all three field components to their quadratic minimum before the experiments, gradients do not change on a week-long timescale.

\section{Correlation function $\gtwo$}
\label{SI:g2}

The normalized second-order correlation function, $\gtwo$ is calculated from the measured coincidence histogram, $G(\tau)$, after subtracting all sources of accidental coincidences arising from background events. The final formula for the signal's correlation is given by:

\begin{equation}
g^{(2)}_{sig}(\tau) = \frac{G(\tau)}{\mathcal{N}(\tau) \cdot N_{cav,A} N_{cav,B}} - \frac{N_{bg,A} N_{bg,B}}{N_{cav,A} N_{cav,B}} - \frac{N_{bg,A}}{N_{cav,A}} - \frac{N_{bg,B}}{N_{cav,B}}
\label{eq:g2bg}
\end{equation}

Here $G(\tau)$ is the raw number of coincidence counts measured in the time bin centered at delay $\tau$, $N_{cav,A/B}$ is the number of signal-only (cavity) counts calculated by subtracting the estimated background counts ($N_{bg}$) from the total measured counts ($N_m$), and $\mathcal{N}(\tau)$ is the normalization factor defined below. The final three terms in \cref{eq:g2bg} subtract the accidental coincidences from background-background, signal-background, and background-signal events, respectively. The normalization factor accounts for the experiment's duration, ensuring $\gtwo$ is dimensionless, and is defined as:
\begin{equation}
\mathcal{N}(\tau) = \frac{N_{shots} \cdot (T_{shot} - |\tau|) \cdot \Delta t}{(N_{shots} \cdot T_{shot})^2}
\end{equation}
where $N_{shots}$ is the number of experimental runs, $T_{shot}$ is the duration of a single shot, and $\Delta t$ is the histogram bin width.

The uncertainty in the final $\gtwo$ is determined by propagating the statistical error from all measured quantities. The primary source of uncertainty is the Poisson counting statistics of the raw event numbers (coincidences, total counts, and background counts), where the variance is equal to the mean number of counts. These individual uncertainties are then combined using standard error propagation formulas for uncorrelated variables to find the final variance on $\gtwo$.

\clearpage
\putbib
\end{bibunit}
%\bibliographystyle{naturemag} % We choose the "plain" reference style
%\bibliography{references}

\end{document}